# Energy-time entanglement from a resonantly driven quantum dot three-level system

M. Hohn,[1] K. Barkemeyer,[2] M. von Helversen,[1] L. Bremer,[1] M. Gschrey,[1] J.-H. Schulze,[1] A. Strittmatter,[1,3] A. Carmele,[2] S. Rodt,[1] S. Bounouar,[1] and S. Reitzenstein[1]

[1]*Institut für Festkörperphysik, Technische Universität Berlin, 10623 Berlin, Germany*

[2]*Institut für Theoretische Physik, Technische Universität Berlin, 10623 Berlin, Germany*

[3]*present address: Institute of Physics, Otto-von-Guericke-University Magdeburg, 39106 Magdeburg, Germany*

*Abstract*

Entanglement is a major resource in advanced quantum technology where it can enable secure exchange of information over large distances. Energy-time entanglement is particularly attractive for its beneficial robustness in fiber-based quantum communication and can be demonstrated in the Franson interferometer. We report on Franson-type interference from a resonantly driven biexciton cascade under continuous wave excitation. Our measurements yield a maximum visibility of (73 ± 2)% surpassing the limit of violation of Bell's inequality (70.7%) by more than one standard deviation. Despite being unable to satisfy a loophole free violation, our work demonstrates promising results concerning future works on such a system. Furthermore, our systematical studies on the impact of driving strength indicate that dephasing mechanisms and deviations from the cascaded emission have major impact on the degree of the measured energy-time entanglement.

*Introduction*

Quantum entanglement is one of the most intriguing predictions in quantum mechanics (QM). Einstein, Podolsky, and Rosen [1] famously asked if QM can be considered complete and motivated intense discussions about the unification of local realism and quantum entanglement. Bell's inequality gave an experimental access to the question if QM can be expanded by a local hidden variable model. His inequality gives a strict limit for measured correlations of entangled particles with a local hidden variable theory [2] and has been tested with high accuracy by numerous experiments [3], including loophole free tests [4–6], that have demonstrated the inability to explain quantum entanglement with a local realist model.

Today, the principle of quantum entanglement has entered various practical fields of applications from quantum computation [7–9] to secure quantum communication [10–12]. One essential task of quantum communication is to deliver a reliable source for entangled photon pairs and enable long-distance distribution, preferably by using the existing telecommunication fiber networks. Many research activities have focused on the generation of polarization-entangled photons with the drawback of polarization mode dispersion [13,14] in optical fibers, which causes decoherence and limits the communication distance. In contrast, energy-time entanglement offers the benefit of stable fiber transmission as demonstrated for a distance up to 10.9 km [15] in a Franson-type setup and 300 km [16] for time-bin entanglement.

A common method to generate energy-time entangled photons is based on spontaneous parametric down-conversion (SPDC), where a nonlinear crystal is pumped by an external laser. One distinguishes energy-time experiments [17,18] where the nonlinear medium is continuously pumped and time-bin experiments [19,20] with a pulsed pump. In the first case, the photon beam is split into pairs of energy-time entangled photons, where the uncertainty of pair emission is ruled by the coherence time of the pump source [21]. The latter method requires a pump interferometer to induce second-order coherence [22]. Although successfully utilized, these sources have the drawback of nondeterministic pair generation, leading to limitations on the accuracy and security in quantum key distribution.

Another attractive option for the generation of entangled photon pairs benefits from the availability of deterministic single-photon emitters. Quantum dots (QDs) show inherently sub-Poissonian statistics and the biexciton-exciton (XX-X) cascade has become a major scheme for the deterministic generation of polarization-entangled photons with high flux [23]. A hurdle for such applications is related to the fine structure splitting (FSS) of QDs which can hinder quantum entanglement by revealing the polarization state of the emitted photons [24,25]. Although successful workarounds like temporal post-selection were developed, the needed technologies are costly and not generally available [26,27]. In contrast, energy-time entanglement is rather insensitive to polarization non-degeneracy and time-bin experiments on semiconductor QDs confirmed the entanglement by quantum state tomography [28-31], and also the generation of multiphoton time-bin entangled states has been reported [32].

Regarding time-energy experiments on semiconductor QDs there is one work under continues off-resonant excitation of a XX-X complex with a Franson visibility of 35% [33], suffering from the incoherent excitation process and another one on the dressed exciton state under coherent resonant excitation yielding a Franson visibility of up to 66%, close to the Clauser, Horne, Shimony and Holt (CHSH)-limit [34] of 70.7% for a violation of Bell's inequality [35]. Here we study the degree of energy-time entanglement from a continuously pumped XX-X cascade in a Franson-type configuration, where the XX state is coherently prepared via resonant two-photon excitation. Former experiments demonstrated the dressing of this three-level system with a large degree of correlation [36]. Based on this excitation scheme, our measurements demonstrate a Franson visibility of up to (73 ± 2)%, which exceeds the CHSH-limit by more than one standard deviation.

*The resonantly driven XX-X cascade*

The system under study consists of a self-assembled InGaAs QD integrated deterministically into a microlens [37]. Figure FIG. 1(a) shows a micro-photoluminescence (µPL) spectrum of the QD-microlens under resonant continuous wave (CW) two-photon excitation. The related excitation scheme is illustrated in Fig. FIG. 1(b). To achieve a cascaded emission, the biexciton state is resonantly pumped by a two-photon process, where the biexciton binding energy of ≈3 meV prevents the simultaneous excitation of the exciton state. The biexciton decays to the ground state via the intermediate horizontally or vertically polarized exciton state, emitting two photons $\gamma_{XX}$ and $\gamma_X$. Polarization dependent measurements reveal a FSS of (28 ± 1) µeV [see inset Fig. FIG. 1(a)] and single-photon emission for the biexciton and exciton state, with a $g^{(2)}_{XX,X}(0)$ close to zero as demonstrated by photon-autocorrelation measurements [Fig. FIG. 1(c,d)]. The cross-correlation function [Fig. FIG. 1(e)] exhibits a bunching with a peak value of 237 normalized coincidences for positive time delays and an anti-bunching for negative time delays, which indicates the time-correlated emission of photons [38]. The exponential decay for positive time delays observed in horizontal polarization (black) is superimposed by oscillations in diagonal polarization (green). This is related to the non-zero FSS of the QD. The energetic splitting causes a phase difference, which can be tracked in the XX-X-correlation in diagonal or antidiagonal base [26,39]. The inset in Fig. FIG. 1(e) shows the measurement for larger time scales and reveals a bunching beneath the XX-X cascade signal. Such a bunching behavior is typically caused by blinking of the emitter which can be attributed to various mechanisms such

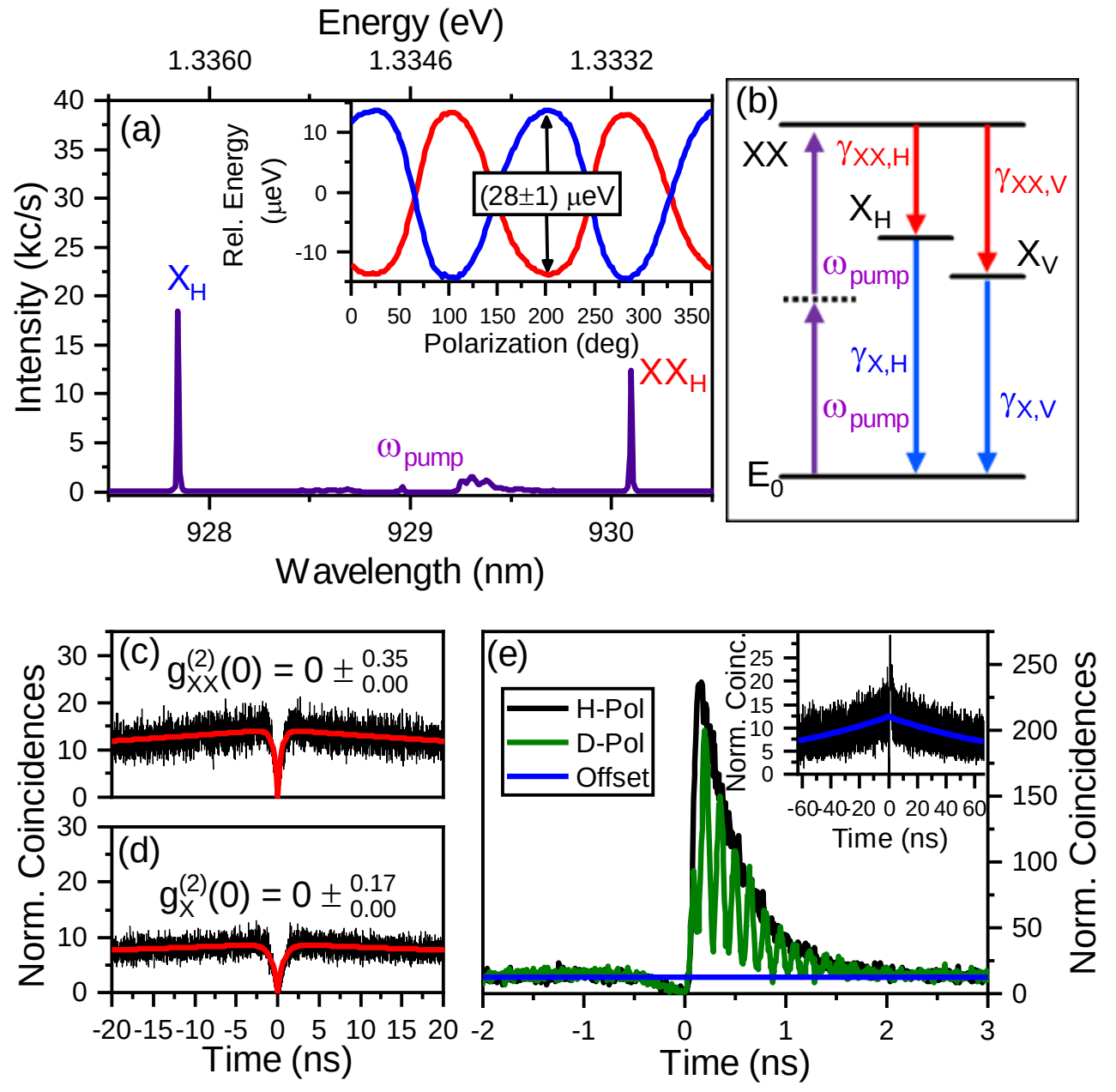

*FIG. 1 (a) µPL spectra in horizontal detection of the QD XX-X cascade. Inset: Relative energy of the X and XX emission lines over polarization angle. (b) Scheme of the two-photon excitation process. (c) Auto-correlation for the XX and (d) X signal and (e) cross-correlation in horizontal and diagonal detection basis. Inset: Bunching-offset on longer time delays approached by an exponential function (blue). All measurements performed at an excitation power of 4.6 µW.*

as spectral diffusion, the occupation of dark states or charging of the QD [40–42]. The level of the bunching offset strongly depends on the excitation power and, as we will discuss later, degrades the degree of energy-time entanglement. However, it is possible to consider the bunching by fitting the offset with an exponential function and subtract it from the data. A consequence of the continues wave resonant excitation is the coupling of the excitonic states with the laser field resulting in new eigenstates. These so-called dressed states have excitation power dependent Eigenvalues which leads to a splitting of the X and XX transitions [43]. Our monochromator has a spectral resolution of approximately 30 µeV and we observe a clear splitting for driving strengths beyond 100 µW but oscillations in our coherence time measurements indicated

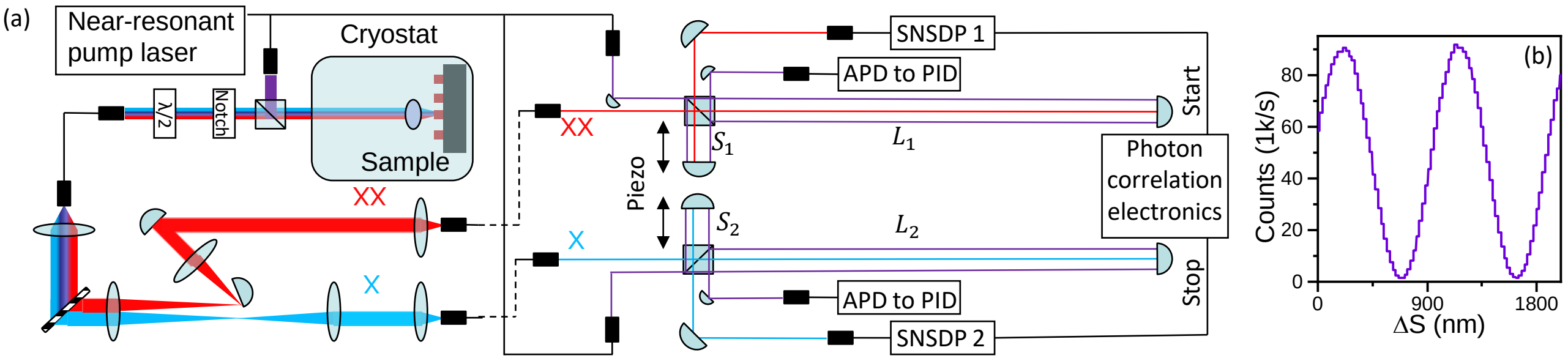

*FIG. 2 (a) Experimental setup. The XX and X signals from a QD are frequency filtered and directed towards two unbalanced interferometers. The phase in each interferometer is adjusted by varying the length of the short arms. For phase stabilization, the pump laser is guided parallel to the signal path and the interference signal (b) is used as reference for a PID controller to compensate deviations in the path length difference over time.*

a splitting already for an excitation power of 15 µW (see Supplemental Material in [44]).

*Franson Interferometer*

In 1989 J. D. Franson developed an idea to realize an experimental test of Bell's inequality for energy-time entangled photons [45]. It consists of an entangled biphoton source and a pair of unbalanced Mach-Zehnder-Interferometers with photon counting devices at the outputs. Additional phase-shift plates ($\phi_i$, $i = 1, 2$) allow one to control the pathlength differences between the short and long arms ($S_i$, $L_i$). Now the pair-wise emitted photons are locally separated and guided in a single interferometer. If the pathlength difference ($\Delta L = L_i - S_i$) is less than the coherence length $l_{coh}$ of the emitted photons, one can observe first-order interference on each detector of the interferometers. Elongating the pathlength difference hinders first-order interference, but a photon pair can still produce interference considering the coincidence detection between the outputs of two interferometers (second-order interference). Such a measurement results in three distinct peaks representing four possible outcomes. The first photon can travel the long, the second the short path ($L_1S_2$) and vice versa ($S_1L_2$). These events can clearly be separated by the temporal path length difference ($\Delta T = (L\text{-}S)/c$) and result in two separated peaks. Events where both photons are detected in the long or short path ($L_1L_2$, $S_1S_2$) cannot be distinguished and consequently result in a single peak at zero-time delay. The photons are in a superposition $|\psi\rangle = 1/\sqrt{2}\,(|S_1,\, S_2\rangle + e^{i(\phi_1+\phi_2)}|L_1,\, L_2\rangle)$, and the coincidence rate interferes for different phase positions $\phi_{1,2}$. Such an interference of distant fields is a manifestation of photon entanglement in time and a visibility beyond the CHSH-limit of 70.7% serves a clear indication of energy time entanglement. Nevertheless, for a genuine violation of Bell's inequality it is necessary to close the "post-selection loophole" [46–48], which arises by discarding events from the side peaks. This was already demonstrated on SPDC sources, by a design adjustment into a hugged interferometer configuration [49].

Utilizing the XX-X cascade as a source of energy-time entangled photons inherits a problem arising from the lifetime ratio of the XX and X state. Theory predicts a drastically reduced visibility for a slower decaying intermediate state as compared to the upper state [50], resulting from a possible emission time leakage of the cascade. In the energy-time entanglement concept presented by Franson, the overall uncertainty of the cascade emission time is creating the energy-time entangled state. Hence, any process reducing this uncertainty reduces the degree of entanglement. We attribute the possibility to violate the CHSH-limit to the continuous wave pump which hinders emission time information leakage. But as a theoretical treatment especially in the dressed state regime is missing and will be subject of future work, we can only assume that the expected Franson visibility originates in an interplay between the coherence time of the pump laser and the lifetime ratio of the XX and X states.

*Results*

The experimental setup used is illustrated in Fig. FIG. 2(a). The sample is excited by a tunable diode laser as a pump source and luminescence is collected by a single aspheric lens with a NA of 0.8. Sample and lens are localized in a closed-cycle cryostat at 4.5 K. Notch filters and a

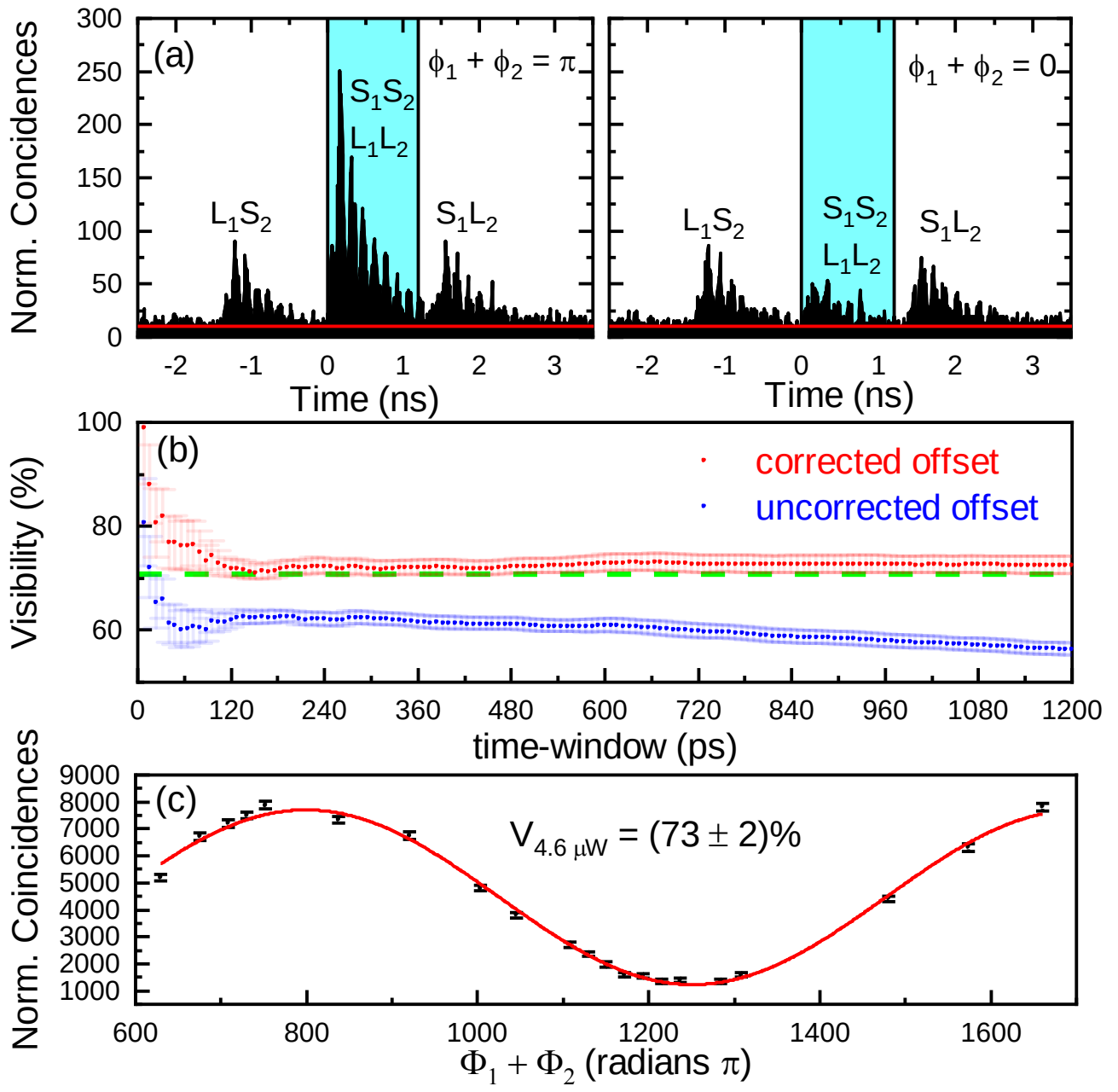

*FIG. 3 (a) Normalized coincidences for maximized (left) and minimized (right) phase settings. The blue shaded area is the integration time window, and the red line indicates the blinking-offset. (b) Visibility over time window for blinking corrected (red) and uncorrected (blue) offset. The green dashed line addresses the CHSH limit. (c) Central peak coincidences and (Poisson) error bars over phase position for a time-window of 1200 ps with blinking correction. The red line indicates the sinusoidal fit.*

combination of a half-wave plate, and a linear polarizer allows us to suppress the pump laser and choose the detected polarization before the signal is collected in a single-mode fiber. The XX- and X-signals are separated at a transmission grating and coupled into the two unbalanced interferometers. Each interferometer consists of a single beam-splitter and two retroreflectors. The pathlengths of the long and short arms are $L = 25$ cm and $S = 3.5$ cm, resulting in a path length difference of $\Delta L = 2\,(L - S) = 43$ cm ($\Delta T$ = 1.4 ns). We observe no one-photon interference because the path length difference surpasses the coherence lengths of the photons ($l_{coh} \approx 14$ cm, determined by Michelson-interferometry) by a factor of three. By adjusting the mirror positions in the short arms with a piezo electric stage, we vary the relative phase. The output of each interferometer is detected with a superconducting nanowire single-photon detector (SNSPD) system, with a combined time resolution of ≈100 ps and detection efficiencies of

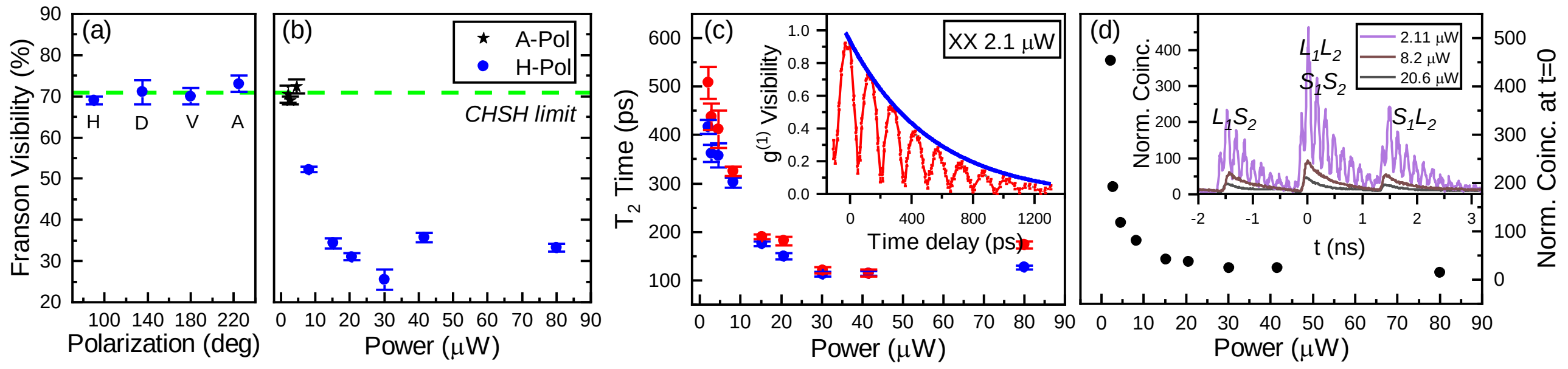

*FIG. 4 Franson visibility over detected polarization (a) and excitation power (b) evaluated by post selection of the central peak and blinking correction. The green dashed lines refer to the CHSH limit. (c) Measured coherence time for the XX (red) and X (blue) signal over excitation power with the Michelson interference of the XX for an excitation power of 2.1 µW (inset). (d) Normalized coincidences at zero-time delay of the summed-up correlations in the Franson interferometer (inset) over excitation power.*

≈85%. To provide a stable phase during measurements the signal of the pump laser is guided parallel to the QD signal. With a coherence length of several km a clear interference signal and first-order interference visibilities of up to 97% are detected via two single-photon counting modules (SPCMs) based on avalanche photodiodes (APDs) [Fig. FIG. 2(b)]. This signal is used as a reference for a PID controller to actively compensate for deviations in the path length difference over time by adjusting the piezo mirror positions in the short arms.

The coincidence detection at the interferometer outputs undergoes Franson interference with varying phase positions. Figure FIG. 3(a) illustrates measurements for a maximized and minimized interference in indistinguishable events $(L_1L_2, S_1S_2)$ of the central peak. For an excitation power of 4.6 µW we detected a count rate of ≈10 kcounts/s in each channel. Measurement time and bin width were 600 s and 8 ps at each of the 21 measured phase positions. The side peak events $(L_1S_2, S_1L_2)$ can clearly be distinguished by the time difference of $\Delta T = \Delta L/c$ = 1.4 ns and stay at constant heights. These measurements were performed in antidiagonal polarization. Corresponding to the cross-correlation measurements in Fig. FIG. 1(e) the decay exhibits oscillations because of the non-zero FSS. The fitted blinking-offset is about 10 normalized coincidences and marked by a red line. As these events originate in interruptions of the cascade, they do not experience any Franson interference. All correlation measurements are normalized with a factor $N = R_1R_2TW$, with count rates $R_1$ and $R_2$, total integration time $T$ and the histogram bin width $W$. From the blue shaded time window between 0 ps and 1200 ps, the number of coincidences is recorded and assigned to the relative phase of the interferometers, as shown in Fig. FIG. 3(c). Afterwards the visibility is extracted from a sinusoidal fit weighted with the Poisson error of the summed-up coincidences. We perform two steps of post selection in the data. First, we select events of the central peak by choosing a time window between 8 ps and 1200 ps. Second, we filter out events not part of the XX-X cascade by subtracting the blinking-offset estimated by an exponential fit of the bunching on larger time scales [Fig. FIG. 1(e)] and indicated in red in Fig. FIG. 3(a). The error of the exponential fit is included in the error of the summed-up coincidences by error propagation. The dependence of the visibility for both processes is illustrated in Fig. FIG. 3(b), with the calculated visibility over time-window for an uncorrected offset, at the accidental level of one normalized coincidence, and the blinking corrected offset. For the uncorrected data (blue) the visibility strongly decays from 82% to 56%. Such a steady decline in visibility is expected for a constant noninterfering background level above the accidental level. With increasing time-window the signal-to-background ratio decreases, because of the exciton decay for positive time delays, which leads to a reduction in visibility. Subtracting the non-interfering background (red) on the other hand results in a nearly constant visibility. For a time-window of 1200 ps, the calculated visibility is V = (73 ± 2)%, which was the maximum achieved visibility in our experiments and is beyond the CHSH inequality by more than one standard deviation.

To obtain more insight into the underlying physics and in interest for an optimized Franson visibility we performed measurements at different polarizations and driving strengths. Here we compare the maximum achieved visibility after a time window of 1200 ps with blinking correction. Measurements at 4.6 µW for horizontal [V = (69 ± 1)%, 12 kcounts/s], vertical [V = (70 ± 2)%, 12 kcounts/s], diagonal [V = (71 ± 3)%, 11 kcounts/s] and antidiagonal [V = (73 ± 2)%, 10 kcounts/s] detection [Fig. FIG. 4(a)] show an improved visibility in diagonal and antidiagonal detection by approx. 3%. A polarization dependence in the Franson interferometer is not expected as long the polarization does not allow to distinguish between the events $L_1L_2$ and $S_1S_2$ [51]. A different factor might be attributed to the difficulty to control the excitation efficiency for these measurements. As we tried to keep the excitation power at 4.6 µW, slight deviations due to changes in the QD environment between the measurements cannot be excluded. Comparing the mean count rate with the visibilities, we find that a high count rate is associated with a low visibility. This could also indicate a power related effect. The impact of the excitation power on the visibility is depicted in Fig. FIG. 4(b). The detected polarization was in antidiagonal direction for excitation powers between 2.1 µW and 4.6 µW and in horizontal for higher powers. The visibility peaks at V = (73 ± 2)% for 4.6 µW and declines to V ≈ 35% for excitation powers exceeding 15 µW. In total, the data is close to an exponential decay. The following mechanisms could hinder a high Franson visibility. First pure dephasing, which originates in the solid-state environment due to phonon interaction and spectral diffusion. Such interactions could reduce the Franson visibility, by the possibility to gain information about the emission time of the cascade from the lattice environment [52]. Furthermore, deviations from the ideal XX-X decay, like single excitation of the excitonic state, re-excitation into the biexcitonic state, or decays into dark states. Such deviations form an incoherent background which manifests partly in the blinking background and can be filtered in the post selection process. Access to the degree of pure dephasing can be achieved by measuring the lifetime ($T_1$) and coherence time ($T_2$) in the XX-X-system. The dephasing time ($T_2^*$) is then given by $1/T_2 = 1/2T_1 + 1/T_2^*$. Under pulsed off-resonant (890 nm) excitation we measured lifetimes of (711 ± 1) ps and (440 ± 20) ps for the exciton and biexciton state. The coherence time was detected in a separate Michelson interferometer under the same conditions as the Franson measurements (see Supplemental Material in [44]). The inset in Fig. FIG. 4(b) shows a Michelson measurement of the XX state at an excitation power of 2.1 µW. With a coherence time of (508 ± 33) ps, below the Fourier limit of twice the lifetime, the effect of pure dephasing is not neglectable. The rapid loss of coherence with increasing driving strength [Fig. FIG. 4(b)] can be attributed to excitation induced dephasing [53,54] and additional non-resonant excitation from the residual straylight of the resonant diode laser. Experimental insight to deviations from the ideal XX-X cascade can be gained from a cross-correlation measurement. The summed up correlation data from the Franson interferometer for different powers (inset **Fehler! Verweisquelle konnte nicht gefunden werden.**(c)) represents the mean probability for the indistinguishable ($L_1L_2$^$S_1S_2$) and distinguishable events ($L_1S_2$, $S_1L_2$), and comparing the normalized coincidences at zero time delay serves an indicator for the excitation power dependent degree of correlation. Figure FIG. 4(c) displays the extracted values obtained from the summed up Franson measurements for driving strengths between 2.1 µW and 80 µW. Comparing the Franson visibility with the decay in coherence time and normalized coincidences at zero-time delay [Fig. FIG. 4(a,b,c)] suggests a clear impact of these parameters on the degree of energy-time entanglement and such measurements can serve as preliminary examination to estimate the degree of energy time entanglement.

Nevertheless, a raise in visibility for excitation powers below 4.6 µW is not detected despite the significant raise in coherence time and normalized coincidences. A possible explanation could be given by the expanded measurement time (20 mins at total 15 phase positions and a count rate of ≈2 kcounts/s) to achieve a sufficient statistic. Such measurements are demanding with respect to the long-term stability of the system and the emitter. As the phase and statistical errors of the correlation measurements are considered in the evaluation process, certain deviations in the QD emitter over integration time could not be completely tracked. In the latter case we observed a decrease in count rate over time, which we attribute to a spectral detuning of the two-photon resonance. This may explain the lowered visibility. Such effects are also not represented in the coherence time measurements as these are taken in a few minutes compared to the Franson visibility measurement over a time of several hours.

*Conclusion*

We investigated the degree of energy-time entanglement in the XX-X cascade of an InGaAs QD under resonant continuous pumping. The measured Franson visibility of (73±2)% is above the CHSH-limit by more than one standard deviation and according to our knowledge the highest value measured in a QD system until today (see Supplemental Material in [44]). We attribute this achievement to the continues coherent excitation regime together with the use of deterministically fabricated QD-microlenses with enhanced photon extraction efficiency and highly efficient SNSPDs, which results in a fairly high count rate at very low driving strengths. We performed two steps of post selection, first by filtering only indistinguishable events of the central peak in the correlation measurements and second by subtracting the blinking events. Our excitation power dependent visibility, coherence time, and correlation measurements indicate that environment-induced dephasing and deviations from the ideal cascaded emission have major impact on the achievable degree of energy-time entanglement. Integration of the QDs into cavities and benefiting from enhanced light-matter interaction could overcome these limitations. Firstly, by an improved extraction efficiency, we can achieve similar count-rates at lower driving strength, which reduces the probability for solitude excitations of the exciton state and re-excitation into the biexciton state. Furthermore, the power related dephasing would be addressed and the enhanced radiative emission by the Purcell effect, could serve a further reduction of dephasing effects [55]. For a genuine violation of Bell's inequality with a QD-system, it is possible to close the "post-selection loophole" with a hugged interferometer setup, as demonstrated for SPDC sources [56] [18].

*Acknowledgement*

The research leading to these results has received funding from the European Research Council (ERC) under the European Union's Seventh Framework ERC Grant Agreement No. 615613, from the German Research Foundation (DFG) via Project No. RE2974/18–1 and CRC 787, and via the SEQUME project (20FUN05) from the EMPIR program co-financed by the Participating States and from the European Union's Horizon 2020 research and innovation program. Finally, we would like to thank Andreas Knorr for his assistance in the discussion of theoretical issues.

# Supplemental material:

## Energy-time entanglement from a resonantly driven quantum dot three-level system

M. Hohn,[1] K. Barkemeyer,[2] M. von Helversen,[1] L. Bremer,[1] M. Gschrey,[1] J.-H. Schulze,[1] A. Strittmatter,[1,3] A. Carmele,[2] S. Rodt,[1] S. Bounouar,[1] and S. Reitzenstein[1]

[1]*Institut für Festkörperphysik, Technische Universität Berlin, 10623 Berlin, Germany*

[2]*Institut für Theoretische Physik, Technische Universität Berlin, 10623 Berlin, Germany*

[3]*present address: Institute of Physics, Otto-von-Guericke-University Magdeburg, Magdeburg 39106, Germany*

This document gives supporting information on the investigated three-level system. Especially we give further insights in the dressed three-level regime, a detailed description of the performed coherence time measurements and of the stabilization procedure.

*Preliminary Investigation*

Prior to the measurements shown in the main text we performed preliminary measurements to identify excitonic (X) and biexcitonic (XX) spectral lines of the studied QD. In a first step we perform off resonant excitation (890 nm) via a 2 ps pulsed tunable optical parametric oscillator (OPO) laser with a repetition frequency of 80 MHz. Figure S**Fehler! Verweisquelle konnte nicht gefunden werden.**(a) illustrates the µPL spectra at an excitation power of 2 µW, with the X state located at 927.9 nm and the XX state at 930.1 nm. We can identify the states by the power dependent intensity [Fig. S**Fehler! Verweisquelle konnte nicht gefunden werden.**(b)]. The intensity in double logarithmic presentation increases with a slope of $m_X$ = 1.02 ± 0.02 for the X state. The XX state starts to emit at higher powers, where the population rate to the s-shell of the QD exceeds the relaxation rate to build up XX complexes. The linear fit yields a slope $m_{XX}$ = 1.68 ± 0.01. The slope of the intensity-power relation is closely affiliated to the relaxation rate and hence the lifetimes of the forementioned states. As mentioned in the main text, the lifetime relation in the XX-X-cascade is expected to be of importance for the degree of energy time entanglement. Furthermore, it is a needed parameter to evaluate the dephasing time from the performed coherence time measurements. We performed time-correlated single photon counting (TCSPC) measurements by filtering the X and XX signals in a transmission spectrometer and detection via superconducting nanowire single-photon detectors (SNSPDs) with a combined time resolution of around 100 ps. The time resolved measurement is illustrated in Fig. S**Fehler! Verweisquelle konnte nicht gefunden werden.**(c). We determine lifetimes of $T_{1,X}$ = (711 ± 1) ps for the X state and $T_{1,XX}$ = (440 ± 20) ps for the XX state. The close relation between the intensity-power slope and the decay rates is demonstrated by the similar quotients $T_{1,X}/T_{1,XX}$ = 1.64 ± 0.08 and $m_{XX}/m_X$ = 1.65 ± 0.04.

*Resonant excitation and dressing of the three-level system*

For resonant excitation we take advantage of the two-photon excitation as described in the main text. The QD is continuously pumped by a narrow external cavity tuneable laser in resonance to the virtual state of the two-photon resonance. A unique feature for resonant driving is the coupling of the excitonic states with the laser field which leads to new eigenstates $|+\rangle, |-\rangle, |0\rangle$ [1,2], referred to as dressed states. Figure S**Fehler! Verweisquelle konnte nicht gefunden werden.**(a) shows the transition of the excitation scheme into dressed states for strong driving and the power dependent spectra is illustrated in Fig. S**Fehler! Verweisquelle konnte nicht gefunden werden.**(b). The exciting pump laser is vertically polarized and leaves the horizontal excitonic state in its bare form. The new eigenvalues are

$$E_0 = 0,$$
$$E_- = -\frac{1}{4}\left(\Delta E_B - \sqrt{\Delta E_B^2 + 8(\Omega\hbar)^2}\right),$$
$$E_+ = -\frac{1}{4}\left(\Delta E_B + \sqrt{\Delta E_B^2 + 8(\Omega\hbar)^2}\right),$$

with the biexciton binding energy $\Delta E_B$ and the Rabi frequency $\Omega$. The corresponding eigenvectors are

$$
\begin{aligned}
|0\rangle &= \frac{1}{\sqrt{2}}(|XX\rangle - |E_0\rangle), \\
|+\rangle &= \frac{1}{\sqrt{2+4\left(\frac{E_-}{\Omega\hbar}\right)^2}}\left(|E_0\rangle + \frac{2E_-}{\Omega\hbar}|X_V\rangle + |XX\rangle\right), \\
|-\rangle &= \frac{1}{\sqrt{2+4\left(\frac{E_+}{\Omega\hbar}\right)^2}}\left(|G\rangle + \frac{2E_+}{\Omega\hbar}|X_V\rangle + |XX\rangle\right).
\end{aligned}
$$

The eigenvalues $E_-$ and $E_+$ are functions of the Rabi frequency and shift with increasing power, resulting in a splitting of the X and XX transition into $R_{+/0}$ and $L_{+/0}$. The depicted $R_-$ and $L_-$ transitions are very close to the two-photon resonance and hence suppressed by the notch filter and residual stray light from the pump source. Our setup is limited to a resolution of approximately 30 µeV and we observe a splitting for driving strengths beyond 100 µW. Nevertheless, we assume a splitting to be present even at lower excitation powers. At least our coherence time measurements reveal oscillations in the Michelson interference visibility occurring for 15 µW [Fig. SFIG. S3(d)].

*Coherence time measurements*

In the main material we show a close relation between energy-time entanglement and coherence time of the emitted photon pairs. Our coherence time measurements are realized in a Michelson interferometer with two adjustable retroreflectors [Fig. S**Fehler! Verweisquelle konnte nicht gefunden werden.**(a)]. The first reflector is mounted on a mechanical stage, allowing for a translation ($\Delta x$) up to 300 cm. The second is mounted on a piezo stage ($\delta x$) and used to scan in a range of several µm. In this way the first-order ($g^{(1)}$) visibility is detected for delays between both arms, in the range of -16 mm to 196 mm (time delay: -107 ps to 1294 ps). The interfering signal is then recorded in a monochromator. In postprocessing the desired spectral parts are fitted with a gaussian function, and the visibility is assigned to the time delay. We performed such measurements at different excitation powers and polarizations corresponding to the energy-time entanglement measurements in the Franson interferometer. Figure S**Fehler! Verweisquelle konnte nicht gefunden werden.**(b) shows the evaluated $g^{(1)}$ visibility of the XX line for an excitation power of 2.11 µW in antidiagonal polarization. The first order autocorrelation is connected to the spectral density of the electric field via a Fourier transformation described in the Wiener-Khinchin theorem [3]. Hence, considering the nonzero fine structure splitting in the power spectrum for a measurement in antidiagonal polarization, we exhibit oscillations enveloped by the exponential coherence decay. In contrast a measurement in horizontal detection gives a solely exponential decay [Fig. S**Fehler! Verweisquelle konnte nicht gefunden werden.**(b)]. We evaluated the coherence time ($T_2$) by an exponential fit for positive time delays. In case of underlying oscillations, we simply excluded such data points from the fit. For an excitation power of 15 µW and 80 µW we see oscillations in the first-order interference, although we measured in horizontal polarization. This may stem from the dressed states, which cannot yet be resolved by our spectrometer but gradually appear for increasing driving strengths [Fig. S**Fehler! Verweisquelle konnte nicht gefunden werden.**(b)].

Active phase stabilization

For active stabilization of the two interferometers, we guide the diode pump laser on a parallel path to the QD signal and use the interference signal as feedback for a PID controller. The controller is than adjusting the voltage on a piezo mount attached with the retro reflector of the short arm. Before starting a measurement, we adjust the stabilization by tuning the piezo voltage over a few periods of the sinusoidal interference fringes. Knowing the wavelength of the pump laser allows us to attribute a certain amplitude to a phase position, where the zero point is arbitrarily defined by the first measured phase point.

To put the stabilization to a test, we guided the diode laser simultaneously into the signal and stabilization path and tracked the phase for several hours. Figure S**Fehler! Verweisquelle konnte nicht gefunden werden.** illustrates the results of such a test on one of the two interferometers. The interference fringes of the signal and stabilization path are shown in Fig. S**Fehler! Verweisquelle konnte nicht gefunden werden.**(a) and the stabilization test on a time scale of 100 min is given in Fig. S**Fehler! Verweisquelle konnte nicht gefunden werden.**(b). For this test measurement we used two power meters to detect the interference signals. In the experiment we switched to avalanche photodiodes (APDs). The phase proved stable with a deviation ruled by the power detection noise. As the sinusoidal interference signal is less sensitive to phase fluctuations near to the extremum and the detection noise is enhanced for higher power, the highest deviations in our stabilization are given close to the maxima of the stabilization path signal. For the final evaluation of the phase error, we determine the standard deviation in the measured fluctuations of the stabilization paths for each single measurement. Although during measurement we cannot determine the phase position in the signal paths directly, Fig. S**Fehler! Verweisquelle konnte nicht gefunden werden.**(b) demonstrates that deviations between signal and stabilization path are reasonable small to justify such procedure.

Comparison to previous reports

There are several measurements that can indicate or quantify entanglement. This work studies the feasibility for a violation of Bell's inequality under continues coherent excitation of the XX-X cascade. Therefor we compare the achieved visibility to previous reports of time-energy and time-bin entanglement. A systematical comparison of the entanglement quality is hindered by the fact that our setup cannot deliver a reconstruction of the density operator by quantum tomography. Nevertheless, we can compare the achieved Franson visibility to the energy basis visibility in a time-bin setup. Table TABLE I summarizes recent results itemized by source and method.

*TABLE I Comparison between reports of time-bin and energy-time entanglement for QD and SPDC sources.*

| Reference | Source | Method | Energy basis visibility | Fidelity | Concurrence |
|---|---|---|---|---|---|
| **[4]** | QD | Time-bin | 67(5)% | 0.88(3) | 0.78(6) |
| **[5]** | QD | Energy-time | 66(4)% | - | - |
| **[6]** | QD | Time-bin | 41.40(3)% | 0.69(3) | 0.41(6) |
| **Our result** | QD | Energy-time | 73(2)% | - | - |
| **[7]** | SPDC | Energy-time | 94.5% | - | - |

## *Figures*

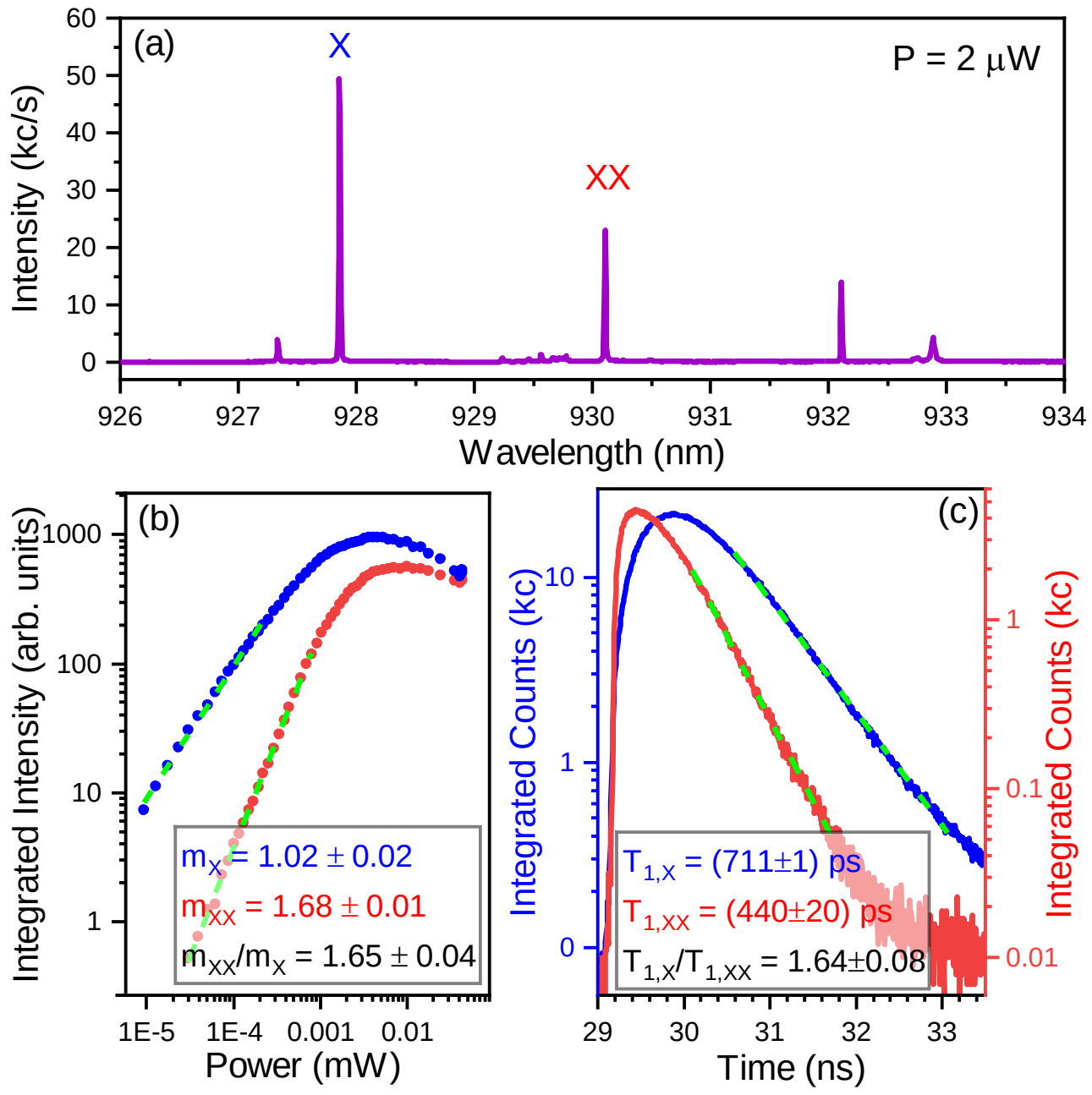

*FIG. S1: (a) µPL spectra under 2 µW pulsed excitation and (b) power dependent integrated intensity for the X (blue) and XX (red) state. Below saturation the power dependent intensity in double logarithmic presentation fits a linear increase (green dashed lines). (c) Lifetime measurement under pulsed off resonant excitation of the X (blue) and XX (red) states. A fit (green dashed line) in the linear region (logarithmic Y-Axes) yields the respective lifetimes.*

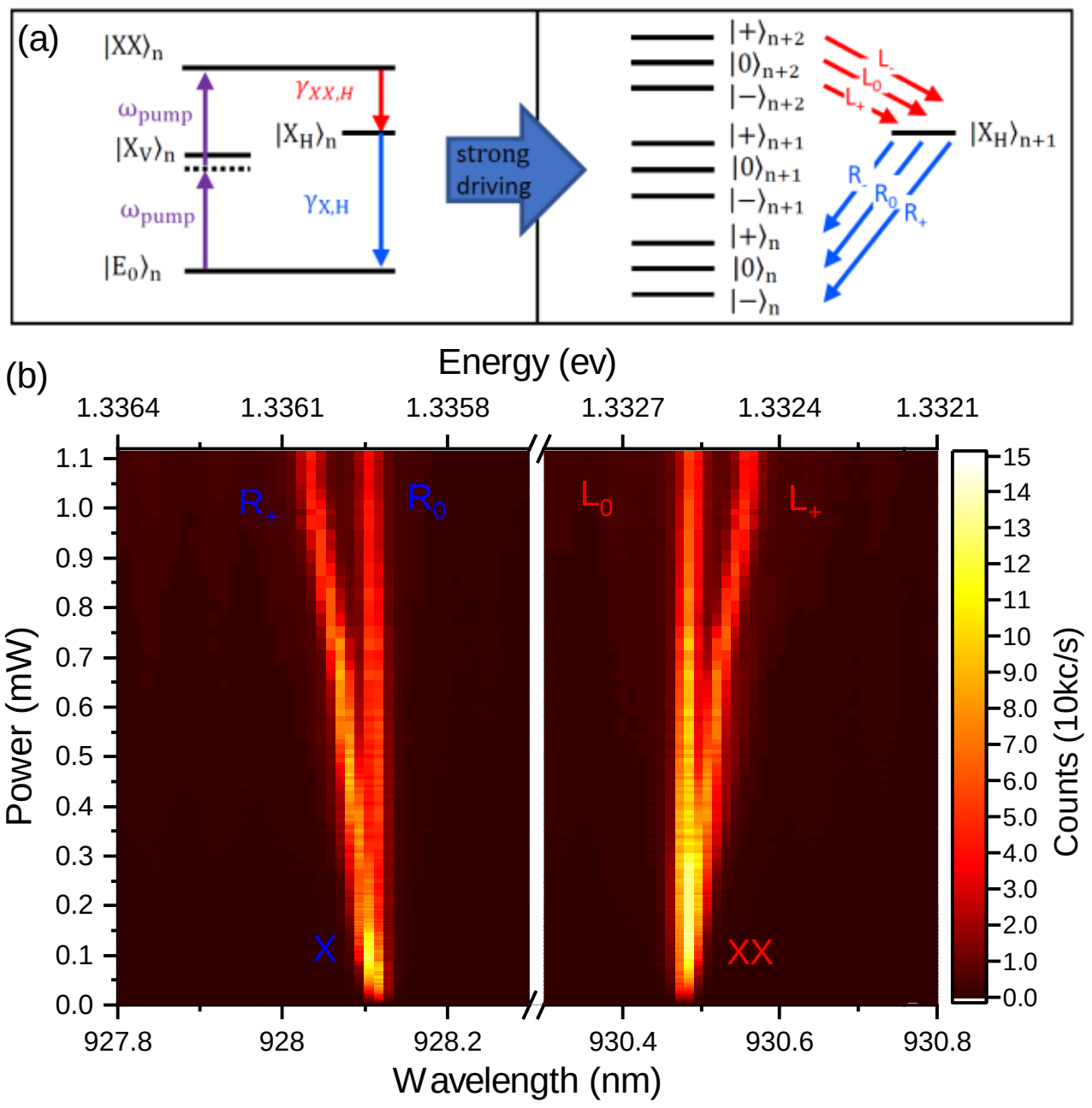

*FIG. S2: (a) Left: Scheme of the two-photon excitation. Right: Dressed states resulting from the coupling of the resonant laser with the excitonic states. (b) Power dependent spectra demonstrating the transition into dressed states $R_+$, $R_0$, $L_+$ and $L_0$.*

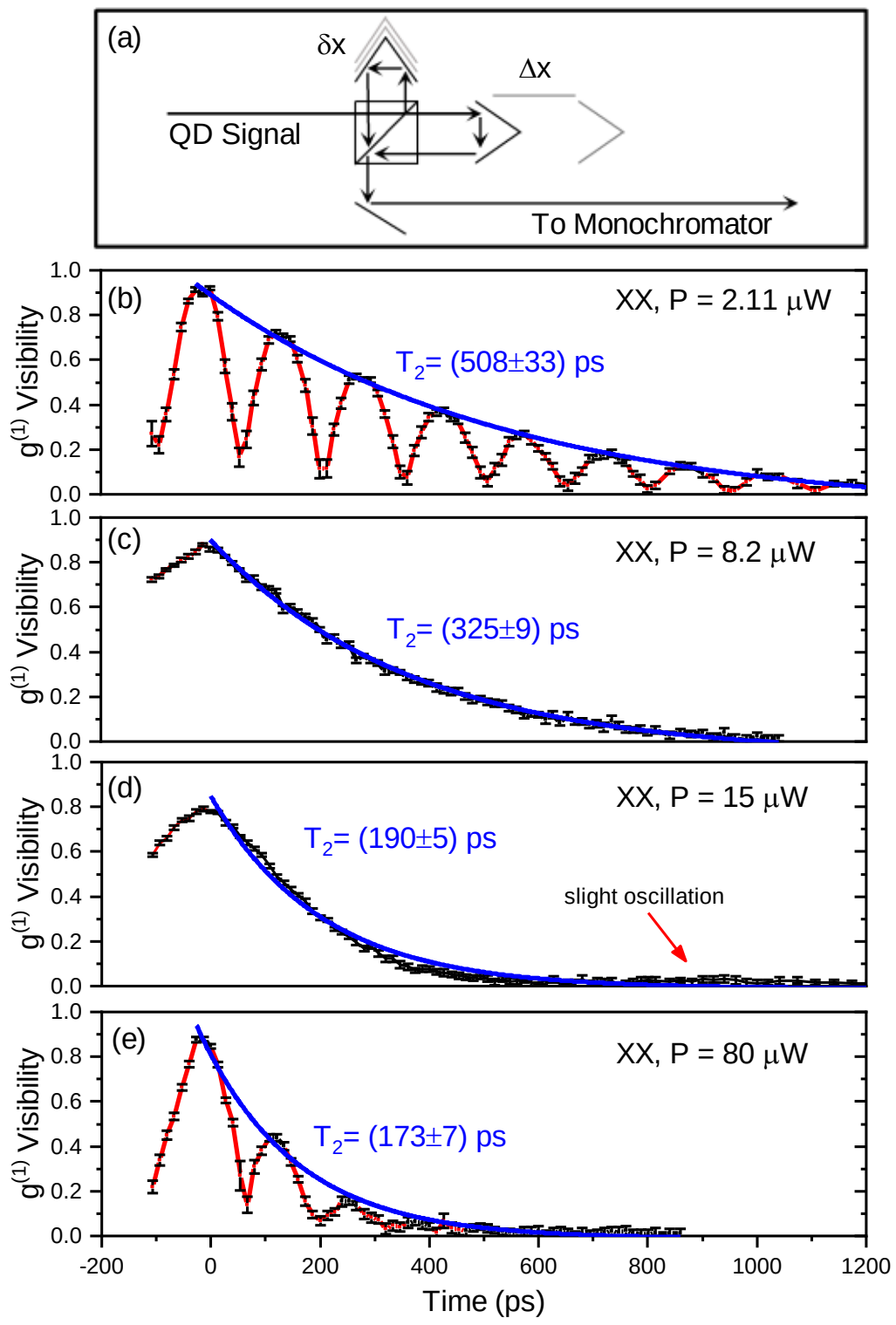

*FIG. S3: (a) Setup of the Michelson interferometer. First-order interference visibility over time delay of the XX signal for a driving strength of 2.11 μW in antidiagonal polarization (b), 8.2 μW (c), 15 μW (d) and 80 μW (e) in horizontal polarization. Red dots are excluded from the exponential fits (blue).*

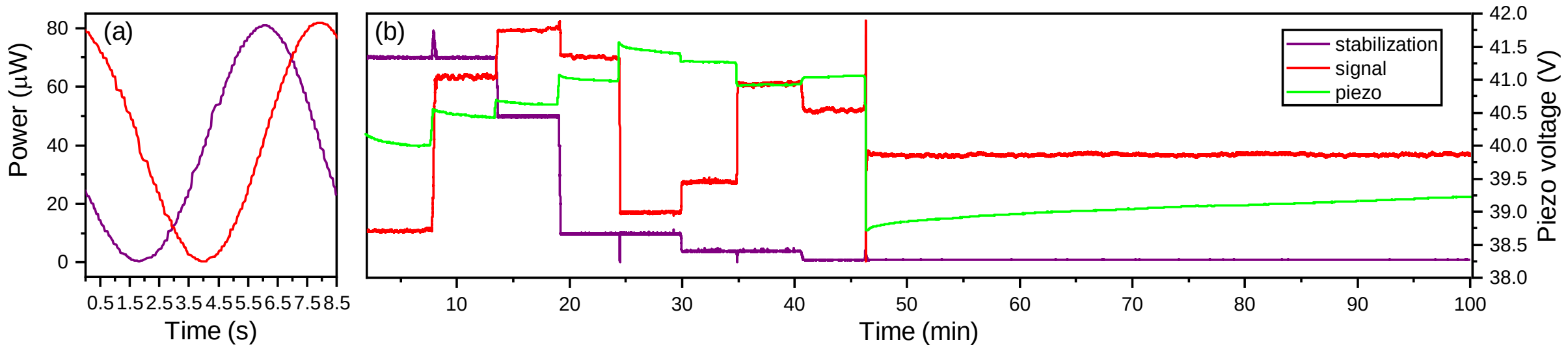

*FIG. S4: (a) Interference fringes of the pump laser guided in the signal and stabilization path of one unbalanced interferometer of the Franson setup. The interference signal is measured with a power meter over time as the short arm is continuously shifted via a piezo mount. (b) Same signal under active feedback of the stabilization to the PID controlled piezo. As the PID controller stabilizes a certain amplitude in the stabilization path (purple), the piezo voltage (green) compensates the drift of the phase also for the signal path (red).*